\documentclass{aa}

\usepackage{graphics}
\usepackage{amsmath}
\usepackage{amssymb}
\usepackage{float}
\usepackage{pslatex}
\usepackage{array}

\newcommand{\cR}{\hbox{$\cal R$}}

\restylefloat{figure}
\restylefloat{table}

\begin{document}


\thesaurus{ 11(11.04.1; 11.12.2) }

\title{ The ESO Slice Project [ESP] galaxy redshift survey:%
\thanks{based on observations collected at the European Southern
Observatory, La Silla, Chile.}}
\subtitle{V. Evidence  for a  D=3 sample dimensionality}

\author{R.Scaramella\inst{1,4}, 
L.Guzzo\inst{2},
G.Zamorani\inst{3,4}, 
E.Zucca\inst{3,4}, 
C.Balkowski\inst{5},
A.Blanchard\inst{6},
A.Cappi\inst{3}, 
V.Cayatte\inst{5},
G.Chincarini\inst{2,12},
C.Collins\inst{7},
A.Fiorani\inst{1}, 
D.Maccagni\inst{8},
H.MacGillivray\inst{9},
S.Maurogordato\inst{10},
R.Merighi\inst{3},
M.Mignoli\inst{3},
D.Proust\inst{5},
M.Ramella\inst{11},
G.M.Stirpe\inst{3}
and G. Vettolani\inst{4}
}

%
\institute{%
Osservatorio di Roma, via dell'Osservatorio 2, 00040 
Monteporzio Catone, Italy
\and {Osservatorio di Brera, via Bianchi 46, 22055 Merate, Italy} 
\and {Osservatorio di Bologna, via Zamboni 33, 40126 Bologna, Italy}
\and {I.R.A./C.N.R. via Gobetti 101, 40129 Bologna, Italy}
\and {Observatoire de Paris, DAEC, Unit\'e associ\'ee 
au CNRS, D0173 et \`a l'Universit\'e Paris 7, 5 Place J.Janssen, 92195
Meudon, France}
\and {Universit\'e L. Pasteur, Observatoire Astronomique, 
11 rue de l'Universit\'e, 67000 Strasbourg, France}
\and {Astrophysics Research Institute, Liverpool 
John--Moores University, Byron Street, Liverpool L3 3AF, United Kingdom}
\and {Istituto di Fisica Cosmica e Tecnologie Relative, 
via Bassini 15, 20133 Milano, Italy}
\and {Royal Observatory Edinburgh, 
Blackford Hill, Edinburgh EH9 3HJ, United Kingdom}
\and {CERGA, Observatoire de la C\^ote d'Azur, 
B.P. 229, 06304 Nice Cedex 4, France}
\and {Osservatorio Astronomico di Trieste, 
via Tiepolo 11, 34131 Trieste, Italy}
\and {Universit\`a degli Studi di Milano, 
via Celoria 16, 20133 Milano, Italy}
}

\date{Received ~~~~~~~~~; Accepted ~~~~~~~~~}

\titlerunning{Evidence for a D=3 sample dimensionality}
\authorrunning{Scaramella et al.}
\offprints{R. Scaramella}

\maketitle


\begin{abstract}

The issue of the approximate isotropy and homogeneity of the
observable universe is one of the major topics in modern Cosmology: 
the common use of the Friedmann--Robertson--Walker [FWR] metric relies
on these assumptions.  Therefore, results conflicting with the ``canonical''
picture would be of the utmost importance.  In a number of recent papers it
has been suggested that strong evidence of a fractal distribution with
dimension $D\simeq2$ exists in several samples, including Abell
clusters [ACO] and galaxies from the ESO Slice Project redshift survey
[ESP].

Here we report  the results of an independent analysis of
the radial density run, $N(<R) \propto R^D$, of the ESP and ACO data.

For the ESP data the situation is such that the explored volume,
albeit reasonably deep, is still influenced by the presence of large
structures.  Moreover, the depth of the ESP survey ($z \la  0.2$)
is such to cause noticeable effects according to different choices of
k-corrections, and this adds some additional uncertainty in the
results.  However, we find that for a variety of volume limited
samples the dimensionality of the ESP sample is $D \approx 3$, and the
value $D = 2$ is always excluded at the level of at least five
(bootstrap) standard deviations. The only way in which we reproduce $D
\approx 2$ is by both unphysically ignoring the galaxy k--correction
and  using Euclidean rather than FRW cosmological distances.

In the cluster case the problems related to the choice 
of metrics and k--correction are much lessened, and we find that
ACO clusters have $D_{ACO} = 3.07 \pm 0.18$ and $D_{ACO} = 2.93 \pm
0.15$ for richness class $\cR \geq 1$ and $\cR \geq 0$,
respectively. Therefore $D=2$ is excluded with high significance also
for the cluster data.



\end{abstract}

\section{Introduction}
Recently Pie\-tro\-ne\-ro and collaborators [hereafter P\&C] (Pie\-tro\-ne\-ro et
al. \cite{Pprinc}, Sylos Labini et al. \cite{SLGMP}, Baryshev et
al. \cite{SLPvistas}) argued that
%
%
a large number of samples give strong evidence that the distribution of
clusters and galaxies is a simple fractal of dimension $D \simeq 2$
\footnote{It must be noticed that this value is different from the
value $D\simeq 1.6$, supported by the same authors in the past years
(Coleman \& Pie\-tro\-ne\-ro \cite{CP}), and $D=2$ has been often
discussed in the literature, because it would naturally arise
\emph{locally} from a geometrical dominance of planar structures, such
as ``pancakes'', see f.i. Guzzo et al. (\cite{gigi+91}).} on
scales of several hundreds of Mpc, quite different from the
``canonical'' value of $D=3$.

This is an important claim because of its far reaching implications,
and this issue, which dates back to the beginning of the century
(Charlier, 1908, see Peebles \cite{peebles}), can be properly
addressed with long and careful analyses which, however, demand much
deeper and better samples than those presently available in order to
be able to explore an adequate range of scales (see e.g. Mc~Cauley
\cite{McCauley97}, Hamburger et al. \cite{Hamburger+96}). 
While 2D constraints come from fluctuations in cosmic backgrounds (see
e.g. Peebles 1993), historically much work on this subject has been
done for decades in analyzing galaxy counts which, in a non evolving
Euclidean Universe, should follow $N(m) \propto 10^{0.6 m}$.  Indeed,
this behaviour has been observed in an intermediate magnitude range,
$m \approx 15 \div 17$, (see e.g. Sandage \cite{sandage}). However, on
the bright side one has to deal with magnitude errors given from
saturation of photographic plates and/or the relatively small volumes
sampled which also reflect in uncertainties in locally the derived
space galaxy density (cf Loveday et al. \cite{SAPM}, Zucca et
al. 1997, Maddox \cite{maddox}), while on the faint side cosmological
curvature and evolutionary effects become dominant and difficult to
disentangle (Koo \& Kron \cite{kookron}, Ellis \cite{ellis}).
Therefore one really needs redshift information in order to avoid
effects of time--space projections.

In this paper we limit ourselves to an
independent check on two of the samples discussed by P\&C, the ESP
galaxies (Vet\-to\-la\-ni et al. \cite{paolo+97}) and the ACO
clusters (Abell, Corwin \& Olowin \cite{ACO}), without touching
upon the very general issue of a possible fractal distribution of the
matter in the universe and its consequences: the interested reader can
consult Peebles (\cite{peebles}), Coleman \& Pie\-tro\-ne\-ro (\cite{CP}), 
Stoeger et al. (\cite{SEH}), Ehlers \& Rindler (\cite{ER}), Szalay \& Schramm 
(\cite{SS}), Luo \& Schramm (\cite{LS}), Mc~Cauley (\cite{McCauley97}), 
Buchert (\cite{Buchert97}), Guzzo
(\cite{gigi97}) and references therein.

However, if the evidence were present in the data at the highly
significant level claimed by P\&C, the simple analysis presented here
should be more than adequate in confirming the $D=2$ claims.

In section 1 we present the formalism, in section 2 the results from
ESP galaxies, in section 3 the results from the ACO clusters, and
finally in section 4 the conclusions.

\section{Method} 
In a previous discussion (Scaramella et al. \cite{io+91}) on the
origin of the Local Group velocity with respect to the Cosmic
Microwave Background frame, it was pointed out the relevance of the
depth's behaviour of clusters' ``monopole'' and ``dipole'' with
respect to the issue of approach to homogeneity of the matter
distribution (see Sylos Labini \cite{Sylos94} for contrasting views).

The method, adopted also by P\&C, is very simple: basically, after
selecting a volume limited sample of objects, one considers the
quantity
\begin{equation}
	N(<R) = \int_0^R dr \, \sum_{i} \delta(r-r_i)     \label{eq:monop}
\end{equation}
where $r_i$ are the radial distances of the objects in the sample and
$\delta$ is the Dirac distribution.
 
Now, for a simple fractal of dimension $D$ one has
\begin{equation}
	\langle N(<R) \rangle = const \cdot R^D		\label{eq:frac}
\end{equation}

One should notice an important difference between the two above
expressions: while Eq.~\ref{eq:monop} refers to a \emph{single point}
(the origin), Eq.~\ref{eq:frac} refers to a statistical property of
the sample, i.e. is obtained by averaging on all points of the sample
(for which several methods have been discussed in the
literature). However, also according to P\&C, for samples with an
adequate number of objects even with Eq.~\ref{eq:monop} one should
recover, after some initial fluctuations (Sylos Labini et
al. \cite{SLGMP}), the ``correct'' dependence on distance of
Eq.~\ref{eq:frac} for a fractal distribution, that is
$	N(<R) = \int_0^R dr \, \sum_{i} \delta(r-r_i) \; \propto \; R^D
$.
%
Therefore P\&C considered an expression equivalent to
Eq.~\ref{eq:monop}, the integral density $N(<R)/R^3 \propto R^{D-3}$,
and derived the value $D \simeq 2$ from preliminary ESP data up to
depths of several hundreds of Mpc.
%
%

\fussy

\section{ESP survey}

The ESP final sample of galaxies (Vettolani et al. 1997, 1998),
limited in apparent magnitude in the $b_J$ band ($b_J \leq 19.4$), is
highly complete in the fraction of measured redshifts ($\sim
85\%$). Despite this fact, its analysis in terms of volume--limited
subsamples is not entirely trivial.

%
\begin{figure}[th]
\resizebox{\hsize}{!}{\includegraphics{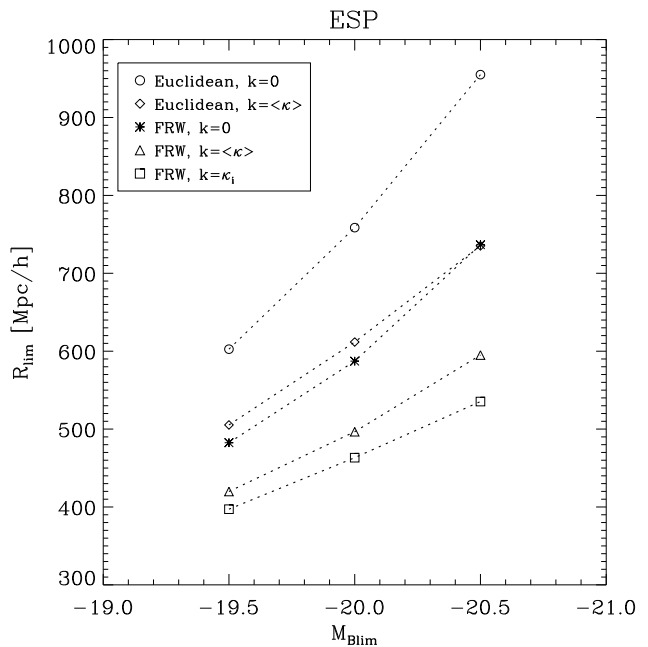}}
\hfill
\caption[distlim.eps]{Limiting distances for ESP volume
limited samples with different
assumptions for metric and k--correction.}
 \label{fig:distlim} 
\end{figure}

The results are somewhat dependent on the assumptions on the
cosmological parameters, and some uncertainty is induced by the
statistical error on the observed magnitudes (r.m.s. of $\simeq 0.2$
mags) and the uncertainties in the adopted k-corrections.

%
%

%
\begin{figure}[ht]
\resizebox{\hsize}{!}{\includegraphics{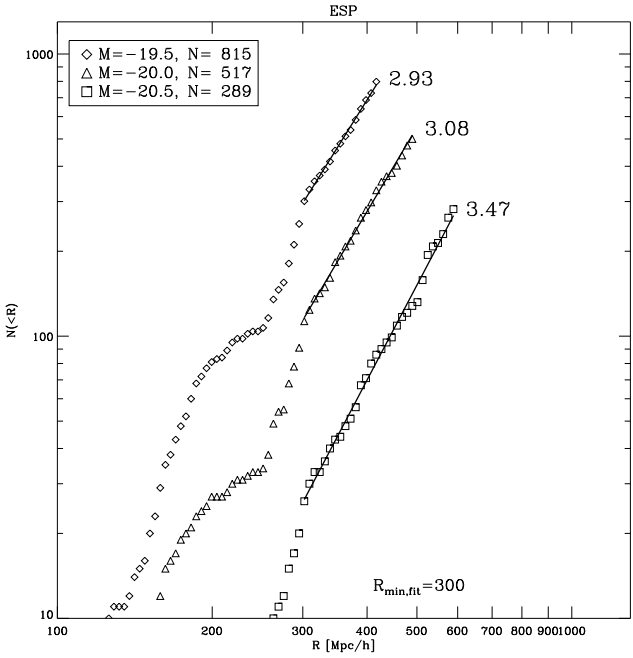}}
\hfill
\caption[avgkcorrfit.eps]{Cumulative depth distribution of 
the number of ESP galaxies for three volume limited subsamples of case
(i), cosmological distance and average k--correction. The value of
slope fitted for $R \geq 300 $ Mpc/h is shown close to the relative
line. The total number of galaxies in each subsample is given in the
legend.}  \label{fig:ESPavgslopes}
\end{figure}

In our analysis we will use the standard formula to derive the absolute
magnitude:
\begin{equation}
M_{b_J} \; = \; b_J - 5 \cdot \left\{ 5 + \log[r_{lum}(z)] + 
\log (h) \right\} -k(z)
\label{eq:B}
\end{equation}
where $k$ is the k--correction term (discussed below), $r_{lum}$ is
the luminosity distance in Mpc (we assume $H_0 =100 $ $ h\,$$ km\,$
$s^{-1}\, $ $Mpc^{-1}$ and $\Omega_0 =1, \Lambda=0$); henceforth we
will drop the $_J$ subscript from the magnitudes.

We will consider ESP volume limited subsamples obtained with three
cuts in absolute luminosity, namely $M_{b_{lim}}= -19.5$, $-20.0$,
$-20.5$. These limits span increasingly deeper volumes with decreasing
statistics.
%
%
Obviously the limiting distances of the volume limited subsamples are
function of the adopted cosmological model and k--correction. In the
following we will consider comoving radial distance: $r_{com} =
r_{lum}/(1+z)$. In the Euclidean case we will have $r_{com} = r_{lum}
= c z/H_0$.

%
%
\begin{figure}[ht]
\resizebox{\hsize}{!}{\includegraphics{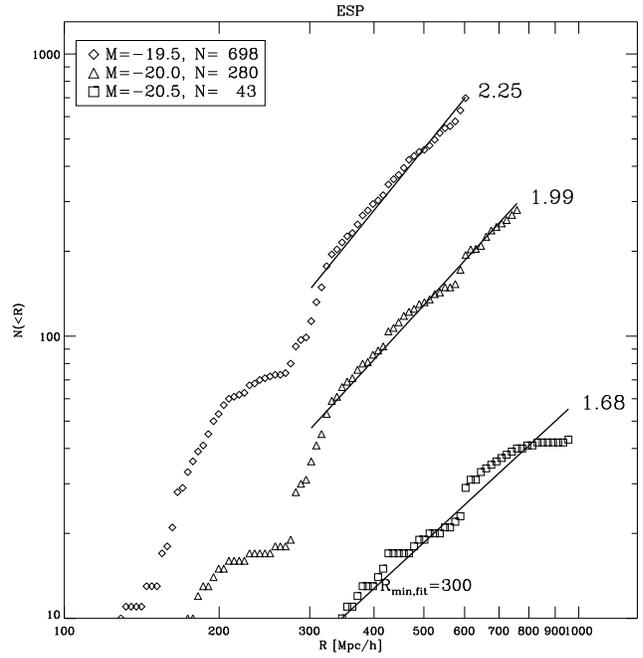}}
\hfill
\caption[euclidzerokcorrfit.eps]{\label{fig:ESPeuclidslopes} 
Cumulative depth distribution of the number of ESP galaxies for three
volume limited subsamples of case (v), Euclidean distance and zero
k--correction.  The value of slope fitted for $R \geq 300 $
Mpc/h is shown close to the relative
line. The total number of galaxies in 
each subsample is given in the legend.}
\end{figure}
%
%

In Fig.~\ref{fig:distlim} we show the limiting distances for
volume--limited subsamples as a function of absolute magnitude for five
different cases: 
\begin{itemize}
\item[(i)] cosmological distance and average
k--correction as in   Zucca et al.   (\cite{zucca+97}); 
\item[(ii)] cosmological distance and  k--correction  estimated
from the spectrum of each galaxy   
(Fiorani \&  Scaramella \cite{ago+io98}); 
\item[(iii)] Euclidean distance and k--correction as in case (i);
\item[(iv)] cosmological distance and zero k--correction; 
\item[(v)] Euclidean distance and zero k--correction.  
\end{itemize}
Except for case (ii) the adopted k--correction at any given redshift
is the same for all galaxies and in these cases the definition of a
volume limited subsample is straightforward. In case (ii) we have to
limit the depth in such a way to include all morphological types, and
therefore we define the limiting distance by using $k_{lim}(z) \simeq 4.1
\, z$, i.e. the k-correction appropriate for elliptical galaxies in
our redshift range, $z \la  0.25$. 

For the ESP sample P\&C claim that the signature of $D=1.9 \pm 0.2$ is
seen for volume limited sub--samples at a depth greater than $300 \,
$~Mpc/h, which according to Sylos Labini et al.  (\cite{SLGMP}) is
the minimum depth of statistical validity of the radial analysis for
this sample.

%
%
We show in Fig.~\ref{fig:ESPavgslopes} the results of fitting $D$ in
Eq.~\ref{eq:frac} to the three volume limited subsamples of case (i)
in the range $R \geq 300 $~Mpc/h.  The influence of the large
inhomogeneity reported in Zucca et al. (\cite{zucca+97}) is evident in
all subsamples up to a depth of $R\sim 300$ Mpc/h, but after that {\it
we find no evidence for a slope $\sim 2$}, as claimed by P\&C.


Since the data presented in Fig.~\ref{fig:ESPavgslopes} are cumulative
distributions, the points are not independent. Therefore in order to
have an estimate of the error associated to the fitted value we
applied the bootstrap method (Efron \& Tbishirani \cite{bootref}),
which yields a measure of the {\it internal uncertainty} of the sample
at hand.  The bootstrap estimates from 10,000 resamplings have a
Gaussian shape and yield: $D_{-19.5}=2.93 \pm 0.14$ , $D_{-20}=3.08
\pm 0.18$ and $D_{-20.5}=3.47 \pm 0.28$.  These values are all
consistent with $D=3$, while for all of them $D=2$ is at more than
five standard deviations.  If we analyze case (ii) (cosmological
distance and k--correction estimated from the spectrum of each galaxy)
we obtain values in good agreement with case (i), as reported in
Table~1.  If we modify case (i) with the use of Euclidean
distances rather than the comoving ones, we still obtain values which
bracket $D=3$ (case iii).

\begin{center}
%
%
%
%
\setlength{\extrarowheight}{2pt}
\begin{tabular*}{\hsize}{@{}rlclll@{}} 
%
%
\multicolumn{6}{l}{ \hbox{Table~1 %
	{\itshape\footnotesize %
	Values of D for volume--limited ESP subsamples}}} \\

\hline
Case & Model & k--correction & M$_{-19.5}$ & 
M$_{-20.0}$ & M$_{-20.5}$ \\
 \hline 
%
%
%
(i) & FRW & $\langle k\rangle$	& 2.93 & 3.08 & 3.47 \\
(ii) & FRW & $k_i$ 		& 2.79 & 3.06 & 3.23 \\
(iii) & Euclid 	& $\langle k\rangle$	& 2.96 & 2.83 & 3.17 \\
(iv) & FRW & $0$		& 2.37 & 2.37 & 2.11 \\
(v) & Euclid & $0$		& 2.25 & 1.99 & 1.68 \\
\hline
%
%
\label{1}  \\
\end{tabular*}
%
%
\end{center}
\medskip

On the contrary, if we neglect the use of k--corrections, we obtain values
closer to $D=2$ (case iv), or even in agreement with $D=2$ (case v:
$D_{-19.5}=2.25 \pm 0.10$ , $D_{-20}=1.99 \pm 0.13$ and
$D_{-20.5}=1.68 \pm 0.26$), thus reproducing the result of P\&C. The
results for the last case are shown in
Fig.~\ref{fig:ESPeuclidslopes}. Since the effect of neglecting the
k--correction term in Eq.~\ref{eq:B}  is to assign
a lower intrinsic luminosity to each galaxy, the brightest subsample
has only $43$ galaxies and is  dominated by
discreteness. 
%
%

The results of the present analysis show the
capital importance of the k--correction term: {\it unless one is
willing to disregard the physical effect of the redshift on the
observed spectrum and intrinsic luminosity} (and moreover to adopt a
purely Euclidean metric), our results do not support the claims of
P\&C of a value  $D=2$ for the ESP survey, but rather suggest the
value $D\simeq 3$. It is worthwhile to stress that the k--correction
is an empirical and necessary correction to the magnitude of a galaxy,
and it does not depend on any interpretation of the redshift. This
correction simply takes into account that at different redshifts we
are looking at different regions of the galaxy spectrum.

\section{Abell clusters}

\begin{figure}[ht]
\resizebox{\hsize}{!}{\includegraphics{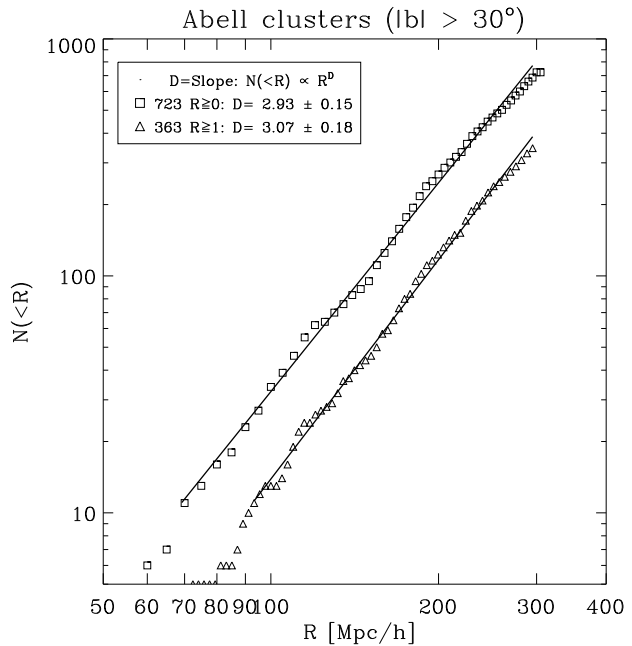}}
\hfill
\caption[acoavgslopes]{Cumulative depth distribution 
of the number of ACO clusters for
volume limited subsamples.}
\label{fig:acoslopes} 
\end{figure}


In agreement with several analyses which appeared on the subject (see
e.g.  Bahcall \cite{bahcrev} and reference therein), Scaramella
et al.  (\cite{io+90}) and Zucca et al.  (\cite{zucca+93})
found that the ACO sample suffers from a significant radial
incompleteness beyond a depth of $300 \, $~Mpc/h.  Even though the
different angular selection functions and overall completeness are
different for the Northern (original Abell) and the Southern sample,
it has been argued (Scaramella et al. \cite{io+90}) that the
radial incompleteness is not very strong within $300 \, $~Mpc/h. In
this case, quite differently from clustering analyses, the angular
selection function should impact only on the amplitude but not on the
radial behaviour of the scalar quantity $N(<R)$.  We will therefore
consider ACO clusters which have $|b| \geq 30^o$ and $R \leq 300
$~Mpc/h without any correction for angular incompleteness.

We show in Fig.~\ref{fig:acoslopes} the results of applying
Eq.~\ref{eq:frac} to the combined North+South sample.  Not all the
clusters have measured redshift: for those for which a measure of $z$
is not available, we have used estimated $z$ from relations which have
$<20\%$ of uncertainty, as reported in the papers above. We have 268
measured $z$ out of 363 for $\cR \geq 1$ and 481 out of 723 for
$\cR \geq 0$.  The clusters with estimated redshifts concentrate
into the $R \ga  250$~Mpc/h region (within which there is $
\sim 85\%$ of measured $z$) and can affect only the last points of
Figure~\ref{fig:acoslopes}

Repeating the same fitting as done in the ESP case, and estimating
similarly the errors on $D$, we obtain $D=3.07\pm 0.18$ and $D=2.93\pm
0.15$ for richness class ${\cal R} \geq 1$ and ${\cal R} \geq 0$,
respectively.

We draw the conclusion that our analysis excludes $D=2$ with high
levels of significance also for clusters, while it favours the
canonical value $D=3$.

\section{Conclusions} 

On the basis of the present analysis we do not find any evidence in the
ESP and ACO samples for a fractal exponent $D\approx 2$ as claimed by
P\&C. For the ESP sample we showed that $D\approx 2$ can be obtained
only by neglecting the k--correction term: in our opinion to neglect
this term would be both unphysical and unjustified. Moreover, also the
cluster sample, which is not significantly affected by this particular
uncertainty, clearly excludes the value $D=2$.

On the contrary, we find evidence that within the errors both samples
show a behaviour on large scales both consistent with and supporting
the canonical value, namely $D=3$. It must be stressed that our
results use direct depth information differently from indirect
arguments such as scaling of angular correlation functions or
fluctuations of cosmic backgrounds (see e.g. Peebles \cite{peebles}).

The above conclusions suggest the need of further and more careful and
sophisticated studies on the claims of a strong evidence for a
fractal matter distribution from the data sets we analyzed. Also
further independent checks should be done and possibly on other,
better suited samples.

\begin{acknowledgements}
We thank  M. Montuori, L.~Pie\-tro\-ne\-ro and F.~Sylos Labini for
instructive and lively discussions on the subject.
\end{acknowledgements}

\end{document}